\documentclass[10pt,  draft, showpacs, amsmath,amsfonts, amssymb,eqsecnum, twocolumn, prd]{revtex4}
\usepackage{youngtab}

\def\a{\alpha}

\def\ep{\varepsilon}

\def\vp{\varphi}

\def\a{\alpha}

\begin{document}
  \title{Skewon no-go theorem}
\author{Yakov Itin}

\affiliation{Institute of Mathematics, The Hebrew University of
  Jerusalem \\ and Jerusalem College of Technology, Jerusalem,
  Israel. \\ email: {\tt itin@math.huji.ac.il}}

\pagestyle{myheadings}
\markboth{Yakov Itin} {Yakov Itin \qquad
Skewon no-go theorem}


 \begin{abstract}
Axion modification of the electrodynamics can be considered as produced by an irreducible part of the constitutive pseudotensor. In this paper, we study the modification of wave propagation produced by the second irreducible part called skewon. We introduce the notions  of skewon optic tensor and skewon optic covector. With these devices we prove that in a pseudo-Riemannian manifold endowed with an arbitrary skewon at least one solution of the dispersion relation  is spacelike. This  means that the skewon generates  superluminal wave motion and is thus  ruled out on the basis of SR principles.  

\end{abstract}
\date{\today}
\maketitle
Although Maxwell electromagnetism is a  theoretically well established and  experimentally confirmed physical theory, its modifications are frequently considered. In particular, axion modified electrodynamics was proposed in \cite{Wilczek:1987mv} and \cite{Carroll:1989vb}. A much richer class of CPT-violation models of electrodynamics was studied in \cite{Kostelecky:2002hh} and \cite{Lammerzahl:2005jw}. 
In \cite{Itin:2004za} we show that axion  can be viewed as an irreducible part of a generic constitutive pseudo-tensor. Although it does not enter the wave propagation in the geometrical optics description it contributes to the higher approximation. 
As it is shown in \cite{Itin:2007wz}, axion yields very hard deformations of the light cone structure and thus can be ruled out on the basis of SR first principles. 
In this paper, we consider the modifications of the light cone structure by the skewon, that is the second irreducible part of the constitutive pseudo-tensor \cite{birkbook}, \cite{NP}, \cite{Obukhov:2004zz}, \cite{Hehl:2005xu}, \cite{LS}. 

In  non-dissipative linear response media, the electromagnetic field can be described by a covariant Maxwell system \cite{birkbook}
 \begin{equation}\label{max}
\epsilon^{ijkl}F_{ij,k}=0\qquad H^{ij}{}_{,j}=J^i\,
\end{equation}
with a linear homogeneous {\it constitutive relation} between the excitation $H^{ij}$ and the field strength $F_{ij}$, 
  \begin{equation}\label{lin-1}
H^{ij}=\frac 12 \chi^{ijkl}F_{kl}\,.
\end{equation}

Due to its definition, the constitutive pseudo-tensor $\chi^{ijkl}$ inherits the symmetries of the skew-symmetric tensors $H^{ij}$ and $F_{ij}$, 
  \begin{equation}\label{lin-2}
\chi^{ijkl}=-\chi^{jikl}=-\chi^{ijlk}\,.
\end{equation}
In four dimensional space,   $\chi^{ijkl}$ has   therefore 36 independent components. 
Under the action of the group $GL(4,\mathbb R)$, such a fourth order tensor possesses a unique irreducible decomposition into the  sum of three independent pieces:
\begin{equation}\label{lin-3}
\chi^{ijkl}={}^{\tt (1)}\chi^{ijkl}+{}^{\tt (2)}\chi^{ijkl}+{}^{\tt
(3)}\chi^{ijkl}\,.
\end{equation}

Here, the third tensor  is totally skew-symmetric, so it has only one independent component. 
Writing it as 
\begin{equation}\label{lin-4}
{}^{\tt (3)}\chi^{ijkl}=\chi^{[ijkl]}=\alpha\varepsilon^{ijkl}\,
 \end{equation}
 the {\it axion pseudo-scalar} $\a$ emerges. This hypothetic co-partner of the photon was intensively discussed in the literature. 

The second tensor of  (\ref{lin-3})  is given by 
\begin{equation}\label{lin-5}   
{}^{\tt (2)}\chi^{ijkl}=\frac
12\left(\chi^{ijkl}-\chi^{klij}\right)\,
 \end{equation}
This {\it skewon  tensor} of 15 components was introduced only recently in  \cite{birkbook}. 

 The first tensor, called the {\it principal part} of $\chi^{ijkl}$, has 20 independent components. It  is expressed as
\begin{eqnarray}\label{lin-7}  
{}^{\tt (1)}\chi^{ijkl}& =&\frac 16\Big(2\left(\chi^{ijkl}+\chi^{klij}\right)-\left(\chi^{iklj}+\chi^{ljik}\right)
\nonumber\\&&-
\left(\chi^{iljk}+\chi^{jkil}\right)\Big)\,.
\end{eqnarray}
In  ordinary Maxwell electrodynamics and its extension to  curved manifolds of GR, the principal part is constructed from the metric tensor
\begin{equation}\label{skew-1}  
{}^{(1)}\chi^{ijkl}=\sqrt{-g}\left(g^{ik}g^{jl}-g^{il}g^{jk}\right)\,.
\end{equation}
In this case, the two other pieces  ${}^{\tt (2)}\chi^{ijkl}$ and ${}^{\tt (3)}\chi^{ijkl}$   are equal to zero. 
 In a non-relativistic anisotropic  linear response material, 
${}^{\tt (1)}\chi^{ijkl} $  decomposes into the standard sets of 3-dimensional electric and magnetic parameters. 

Thus the principle part can be considered as a set of classical parameters aggregated into a covariant 4-dimensional tensor. In  contrast, the axion and skewon tensors do not have classical analogs. Their existence (or non-existence) is an intriguing problem for both theoretical analysis and experimental investigations. Moreover, if such additional electromagnetic parameters could be implemented  in artificial materials, they might have important applications for modern technology. 

In order to study  light propagation in a generic medium equipped with the whole set of constitutive constants, one applies the geometric optic approximation \cite{birkbook}, \cite{Perlick:2010ya},  \cite{Itin:2012ann}. 
Here the system of differential equations (\ref{max}) is substituted by the 
the system of 8 algebraic equations
\begin{equation}\label{char-1}  
\ep^{ijkl}q_jf_{kl} =0\,,\qquad\qquad \chi^{ijkl}q_jf_{kl} =0\,,
\end{equation}
where $q_i=\partial \vp/\partial x^i$ is the wave covector corresponding to the wavefront $\vp(x^i)=0$. The components of the tensor $f_{kl}$ are constructed from  the limiting values of the true field strength $F_{kl}$. When the Fourier transforms exist, they turn into the  phasors of $F_{kl}$. 

The first equation of (\ref{char-1}) is easily solved:
\begin{equation}\label{char-2}  
f_{kl} =\frac 12 (a_kq_l-a_lq_k)\,,
\end{equation}
where   $a_k$ is an arbitrary covector, which is an algebraic analog of the electromagnetic potential. 
Substituting (\ref{char-2}) into the second equation of (\ref{char-1}) we obtain the {\it characteristic system}
\begin{equation}\label{char-3}  
{M^{ik}a_k=0\,,}
\end{equation}
where the characteristic matrix is given by 
\begin{equation}\label{char-3x}  
M^{ik}=\chi^{ijkl}q_jq_l\,.
\end{equation}
 Observe the straightforward  identities
\begin{equation}\label{char-3x} 
M^{ik}q_k=0\qquad {\rm and }\qquad M^{ik}q_i=0\,,
\end{equation}
which are  consequences of the gauge invariance of our system. Thus the matrix $M^{ik}$ is  singular. 
The system (\ref{char-3}) has a trivial solution proportional to $q_k$ which does not  contribute to field strength. In order to have a physically  meaningful solution, the condition 
\begin{equation}\label{disp-2}  
{\rm adj}(M)=0\,
\end{equation}
must be applied \cite{Itin:2009aa}. 
 Recall the standard expression of the fourth order adjoint matrix (${\rm adj}(M)=A_{ij}$)
\begin{equation}\label{disp-3}  
A_{ij}=\frac 1{3!}\ep_{ii_1i_2i_3}\ep_{jj_1j_2j_3}M^{i_1j_1}M^{i_2j_2}
M^{i_3j_3} \,.
\end{equation}
Consequently, the dispersion relation (\ref{disp-2}) reads $A_{ij}=0$ or
\begin{equation}\label{disp-3}  
 \ep_{ii_1i_2i_3}\ep_{jj_1j_2j_3}M^{i_1j_1}M^{i_2j_2}
M^{i_3j_3}=0 \,.
\end{equation}
It turns out \cite{Itin:2009aa}, that we have only one independent relation  here. Indeed,  for a singular matrix satisfying (\ref{char-3x}),  the  adjoint  matrix $A_{ij}$ is proportional to the tensor product of $q_i$,
\begin{equation}\label{Adj2}
A_{ij}=\lambda(q)q_iq_j\,.
\end{equation}
Consequently, the covariant dispersion relation  is represented by a scalar equation  
    \begin{equation}\label{disp-5}
\lambda(q)=0\,.
\end{equation}
where $\lambda(q)$ is a 4-th order polynomial in $q$ and a 3-rd order polynomial in $\chi$. 

Wave propagation is  completely defined by the tensor $M^{ik}$, which is 
 an analog of the acoustic (Christoffel) $3\times 3$ matrix and can be referred to as {\it optic tensor}. Substituting the decomposition (\ref{lin-3}) into (\ref{char-3x}), we have
\begin{equation}\label{chis-1}  
M^{ik}={}^{(1)}\chi^{ijkl}q_jq_l+{}^{(2)}\chi^{ijkl}q_jq_l
+{}^{(3)}\chi^{ijkl}q_jq_l\,.
\end{equation}
Observe that because of the skew-symmetry of ${}^{(3)}\chi^{ijkl}$ the last term  is identically zero, so the axion part   does  not contribute to  wave propagation. Indeed the known role of the axion as a possible source of  CPT-violation \cite{Carroll:1989vb} appears only beyond the geometrical optic approximation \cite{Itin:2004za}.
The two remaining terms on the right hand side of (\ref{chis-1}) are  in general non zero.  
Consequently the optic matrix   decomposes as  
\begin{equation}\label{chis-2}  
M^{ik}=P^{ik}+ Q^{ik}\,.
\end{equation}
Here, the {\it principle optic tensor } $P_{ik}$ and the {\it skewon optic tensor }  $Q_{ik}$ are defined by 
\begin{equation}\label{chis-3}  
P^{ik}={}^{(1)}\chi^{ijkl}q_jq_l\,,\qquad Q^{ik}={}^{(2)}\chi^{ijkl}q_jq_l\,.
\end{equation}
It follows from these definitions that these tensors satisfy
\begin{equation}\label{chis-4}  
P^{ik}=P^{ki}\,,\qquad  Q^{ik}=-Q^{ki}\,.
\end{equation}
Moreover, they are both  singular
\begin{equation}\label{chis-5}  
P^{ik}q_k=0\,,\qquad {\rm and}\qquad Q^{ik}q_k=0\,.
\end{equation} 
Since the second equation of (\ref{chis-5}) involves a skew-symmetric matrix, it can be easily solved
\begin{equation}\label{chis-11}  
Q^{ik}=\ep^{ijkl}q_jY_l\,.
\end{equation}
Here the covector   $Y_i$ is a 1-st order polynomial in $q$. It is only defined   up to the   addition of an arbitrary wave covector $q_i$. Thus an  additional gauge fixing condition  can be imposed on  $Y_i$. 

Since skewon part of the constitutive tensor has 15 independent components, it can be expressed as a traceless quadratic matrix. 
Such representation was derived in \cite{birkbook}:
\begin{equation}\label{mat-1}  
^{(2)}\chi^{ijkl}=\ep^{ijmk}S_m{}^l-\ep^{ijml}S_m{}^k\,.
\end{equation}
Here $S_m{}^k$ is a traceless tensor, $S_m{}^m=0$. 
Correspondingly, the skewon optic tensor is represented as 
\begin{equation}\label{mat-2}  
Q^{ik}=\ep^{ijmk}S_m{}^lq_jq_l\,.
\end{equation}
Rearranging the indices we write it as 
\begin{equation}\label{mat-3}  
Q^{ij}=\ep^{ijkl}S_l{}^mq_kq_l\,.
\end{equation}
and compare with the definition of the skewon covector (\ref{chis-11}). Consequently we have
\begin{equation}\label{mat-4}  
Y_i=S_i{}^jq_j+\a q_i\,,
\end{equation}
where $\a$ is an arbitrary zero order homogeneous function of the wave covector $q$. Different choices of the parameter $\a$ represent different gauge conditions. 

Substituting (\ref{chis-2}) and  (\ref{chis-11})  into (\ref{disp-5}) and calculating the adjoint we obtain that the  dispersion relation for a most generic linear constitutive pseudo-tensor  reads
\begin{equation}\label{disp-24}  
\lambda(P)+P^{ij}Y_iY_j=0\,.
\end{equation}

We will study this equation for a medium with a simple principal part generated by a metric (\ref{skew-1}) and with a generic skewon part ${}^{\tt (2)}\chi^{ijkl}$. Substituting (\ref{skew-1}) into (\ref{disp-24}) we derive the following  very simple form of the dispersion relation
\begin{equation}\label{skew-14}
q^4  =Y^2q^2-(Yq)^2\,.
\end{equation}
Note that this equation is preserved when a "gauge transformation" $Y_i\to Y_i+\a q_i$ with an arbitrary $\a$  is applied. So we can write it  in an even  simpler form by applying a gauge condition on $Y_i$. 
 We chose the Lorenz-type gauge condition
\begin{equation}\label{skew-15}
(Yq)=0\,
\end{equation}
and we are left with the dispersion relation
\begin{equation}\label{skew-15}
q^4  =Y^2q^2\,.
\end{equation}
This equation is not as trivial as it first appears. Note that  the covector $Y_i$ is a 1-st order homogeneous function of $q_i$, but it is not necessarily a polynomial. 
 Indeed, in the Lorenz-type gauge (\ref{skew-15}), the optic covector (\ref{mat-4}) is expressed  as 
\begin{equation}\label{skew-16c}
Y_i=S_i{}^jq_j-\frac{S^{mn}q_mq_n}{q^2}q_i\,.
\end{equation}
In general, the second term here is equal to zero.  See, for instance, the spatially isotropic skewon field of Nieves and Pal \cite{NP} with $S_i{}^j\sim {\rm diag}(-3,1,1,1)$. Another explicit examples of such behaviors are exhibited in \cite{Obukhov:2004zz}. Thus (\ref{skew-15}) does not  necessary have a light-like solution  $q^2=0$.

In order to analyze the solutions of (\ref{skew-15}), we first observe that its left hand side is nonegative. Thus in the case $q^2\ne 0$, the "squares" $Y^2$ and $q^2$ have the same sign. The covectors  $Y_i$ and  $q_i$ cannot, however, be both timelike. Indeed, due to the reverse Cauchy--Schwarz inequality  the product of two timelike vectors cannot be equal to zero, see \cite{Giulini:2006uy}. Consequently,  a non-lightlike solution of (\ref{skew-15}) must be spacelike. 
On the other hand, the dispersion relation can have a solution with $q^2=0$. For instance, it appears when the  tensor $S_{ij}$ in (\ref{skew-16c}) is antisymmetric. If the second solution of the dispersion relation is real the model describes birefringence. 
Rewrite (\ref{skew-14}) as 
\begin{equation}\label{skew-14x}
q^2(Y^2-q^2)  =(Yq)^2\,.
\end{equation}
If this dispersion relation has a lightlike solution $q^2=0$, then (\ref{skew-15}) holds and we remain with  
\begin{equation}\label{skew-16}
q^2  =Y^2\,.
\end{equation}
Recall that we are looking for the solutions that satisfy  the gauge  condition $(Yq)=0$.  
Once more, the covectors $Y_i$ and $q_i$ cannot be timelike and orthonormal at the same time. Consequently they must be  spacelike.  
So we have proved the following\\

{\bf Theorem:} {\it  For an arbitrary skewon modification of a pseudo-Riemannian space, the non-null solution of the  dispersion relation is spacelike.}\\

Since the wave covector $q_i=(w,{\bf k})$ must describe a true  (measurable) wave, such a wave would propagate with a velocity higher than the universal speed of light constant. This is in  direct conflict with the special relativity paradigm. 
 In the curved vacuum of GR, the skewon part of the constitutive tensor must consequently be completely ruled out.

Two short remarks should be added to this result:\\

{\it 1. Isotropic material.} The considerations above remain completely unaltered when we deal with an isotropic medium  with two standard parameters $\varepsilon$ and $\mu$. In this case, however, the wave velocity for the skewon modified medium only  exceeds  the value $\sqrt{\varepsilon\mu}$ instead of the universal speed of light. Such behavior is certainly  not excluded by any fundamental principle. Thus the skewon part  could exist in a material as a phenomenological quantity.  It could for instance   be useful for the description of  dissipative media. Moreover, waves with velocity  higher than $\sqrt{\varepsilon\mu}$ could point towards the presence of   the skewon. \\

{\it 2. Euclidean signature.} The considerations above also remain
valid   for a Riemannian manifold with the Euclidean signature. Here the only change will be the leading sine of  the left hand side of Eq. (\ref{skew-14}). 
In this case, however, the ordinary Cauchy--Schwarz inequality   immediately yields $q^2=0$ . Thus the unique solution is $q_i=0$. Consequently any skewon modification of Euclidean space does not allow wave propagation. 

\section*{Acknowledgements}
  I acknowledge  the GIF grant No. 1078-107.14/2009 for financial support.

\end{document}